\documentclass[letterpaper,12pt]{article}
\usepackage{amsmath}
\usepackage{graphicx,psfrag,epsf}
\usepackage{enumerate}
\usepackage{mdframed}
\usepackage{natbib}
\usepackage{textcomp}
\usepackage[hyphens]{url} 
\usepackage{hyperref}
\usepackage{lmodern}
\usepackage[mathscr]{euscript}
\usepackage{color}

\usepackage{amssymb}
\usepackage{comment}
\usepackage{multirow}
\usepackage{tcolorbox}
\usepackage{eurosym}
\usepackage{booktabs}

\newcommand{\method}{{\tt eLPAR}}
\newcommand{\rating}{r}
\newcommand{\xG}{{\tt xG}}
\newcommand{\player}{p}
\newcommand{\attack}{A}
\newcommand{\budget}{\mathcal{B}}
\newcommand{\middlefield}{M}
\newcommand{\defense}{D}
\newcommand{\gk}{GK}
\newcommand{\positions}{\pi}
\newcommand{\linep}{l}
\newcommand{\linepSet}{L}
\newcommand{\dev}{d}
\newcommand{\positionSet}{\Pi}
\newcommand{\formation}{\phi}
\newcommand{\xGC}{{\tt xGC}}
\newcommand{\mv}{v}
\newcommand{\cost}{c}
\newcommand{\wage}{w}
\usepackage{xcolor}
\usepackage{epstopdf}
\usepackage{etoolbox}
\usepackage{tcolorbox}
\usepackage{tabularx}
\usepackage{array}
\usepackage{colortbl}
\tcbuselibrary{skins}
\newcommand{\bs}{B_s}
\DeclareRobustCommand{\officialeuro}{%
  \ifmmode\expandafter\text\fi
  {\fontencoding{U}\fontfamily{eurosym}\selectfont e}}

\newcolumntype{Y}{>{\raggedleft\arraybackslash}X}

\tcbset{tab1/.style={fonttitle=\bfseries\large,fontupper=\footnotesize\sffamily,
colback=yellow!10!white,colframe=red!75!black,colbacktitle=maroon!40!white,
coltitle=black,center title,freelance,frame code={
\foreach \n in {north east,north west,south east,south west}
{\path [fill=red!75!black] (interior.\n) circle (3mm); };},}}

\tcbset{tab2/.style={enhanced,fonttitle=\bfseries,fontupper=\footnotesize\sffamily,
colback=yellow!10!white,colframe=red!50!black,colbacktitle=blue!40!white,
coltitle=black,center title}}

\tcbset{tab3/.style={enhanced,fonttitle=\bfseries,fontupper=\footnotesize\sffamily,
colback=yellow!10!white,colframe=red!50!black,colbacktitle=red!40!white,
coltitle=black,center title}}

\usepackage{mathptmx}

\usepackage{graphics}
\usepackage{amsthm}
\usepackage{amssymb}
\usepackage{bm}

\newcommand{\todonew}[1]{{\color{black} #1 }}

\newcommand{\todo}[1]{{\color{black} #1 }}

\begin{document}

\def\spacingset#1{\renewcommand{\baselinestretch}%
{#1}\small\normalsize} \spacingset{1}

\title{\bf A Skellam Regression Model for Quantifying Positional Value in Soccer}

 \author{ Konstantinos Pelechrinis$^1$, Wayne Winston$^2$}
 
 \date{%
    \footnotesize
     $^1$School of Computing and Information, University of Pittsburgh \\
     $^2$Kelley School of Business, Indiana University  \\
    \today
}

\maketitle

\bigskip

\begin{abstract}
Soccer is undeniably the most popular sport world-wide and everyone from general managers and coaching staff to fans and media are interested in evaluating players' performance. 
Metrics applied successfully in other sports, such as the (adjusted) +/- that allows for division of credit among a basketball team’s players, exhibit several challenges when applied to soccer due to severe co-linearities.  Recently, a number of player evaluation metrics have been developed utilizing optical tracking data, but they are based on proprietary data. 
In this work, our objective is to develop an open framework that can estimate the expected contribution of a soccer player to his team’s winning chances using publicly available data. In particular, using data from (i) approximately 20,000 games from 11 European leagues over 8 seasons, and, (ii) player ratings from the FIFA video game, we estimate through a Skellam regression model the importance of every line 
(attackers, midfielders, defenders and goalkeeping) in winning a soccer game. We consequently translate the model to expected league points added above a replacement player ({\method}). This model can further be used as a guide for allocating a team's salary budget to players based on their expected contributions on the pitch. 
We showcase similar applications using annual salary data from the English Premier League and identify evidence that in our dataset the market appears to under-value defensive line players relative to goalkeepers.
\end{abstract}

\noindent%

\section{Introduction}
\label{sec:intro}

Soccer is undoubtedly the {\em king of sports}, with approximately 4 billion global following (\cite{worldatlas}). 
\todo{While soccer clubs were slower in embracing the world of data and computational insights as compared to other sports, they have taken huge strides during the last few years integrating {\em analytics} in their soccer operations. 
}
Traditional soccer metrics quantify on-ball events (e.g., corners, fouls, shots etc.). 
However, soccer epitomizes the notion of team sports through a game of space and off-ball movement. 
In soccer every player has possession of the ball an average of only 3 minutes (\cite{fernandez2018wide}), and hence, metrics that quantify on-ball events will fail to capture a player's influence on the game. 
Nevertheless, novel optical tracking technology has allowed for data collection and analysis of off-ball events and movements, hence, providing us with a more complete view of a player's contributions.  
Unfortunately, these approaches are not fully open, in the sense that these type of data are not available to the public. 
The objective of this study is to develop a fully reproducible and open method, for evaluating the expected contribution of a player on the pitch. 

One of the most widely used advanced metrics is expected goals ({\xG}) (\cite{lucey2015quality,fairchildspatial}). 
{\xG} takes into account the context of a shot (e.g., location, number of defenders in the vicinity etc.) and provides us with the probability of the shot leading to a goal. 
{\xG} allows us to statistically evaluate players, while also allowing us to evaluate things such as a team's {\em luck} during a game. 
For example, if a player is over-performing his expected goals, it suggests that he is either lucky or an above-average finisher. 
If this over-performance persists year-after-year then the latter will be a very plausible hypothesis. 
Nevertheless, while expected goals represent a straightforward concept and has been already used by mainstream soccer broadcast media, its application on evaluating players is still limited to a specific aspect of the game (i.e., shot taking) and only to players that actually take shots (and potentially goalkeepers). 
A more inclusive version of {\xG}, is the Expected Goal Chains {\xGC} (\cite{xGC}). 
{\xGC} considers all passing sequences that lead to a shot and credits each player involved with the expected goal value for the shot. 
Of course, not all passes are created equally (\cite{Power:2017:PCE:3097983.3098051}) and hence, {\xGC} can over/under estimate the contribution of a pass to the final shot.

As aforementioned, the last few years player tracking technology has started penetrating the soccer industry. 
During the last world cup in Russia, teams obtained player tracking data in real time (\cite{economist-worldcup})! 
The availability of fine-grained spatio-temporal data have allowed researchers to start looking into more detailed ways to evaluate soccer players through their movement in space. 
For example, (\cite{le2017coordinated,le2017data}) developed a deep imitation learning framework for identifying the {\em optimal} locations - i.e., the ones that minimize the probability of conceding a goal - of the defenders in any given situation based on the locations of the attackers (and the other defensive players). 
Furthermore, (\cite{fernandez2018wide}) also analyzed player tracking data and developed a metric quantifying the contribution of players in space creation as well as this space's value, \todo{while (\cite{fernandez2019decomposing}) developed a deep learning model to obtain an expected possession value/goal metric that is able to evaluate the various actions that take place on the pitch. 
Almost at the same time, (\cite{Decroos19}) developed VAEP, a metric for valuating events based on the probability added in terms of scoring. 
They further developed player ratings by summing up the VAEP of all actions taken by a specific player.  
This area is fast-growing and a nice overview of the current status of advanced spatio-temporal soccer analytics is provided by (\cite{doi:10.1111/j.1740-9713.2018.01146.x}). }
Player tracking data undoubtedly provide managers, coaches and players with information that previously was considered to be {\em intangible} revolutionizing soccer analytics. 
However, one of the problems with these data is their availability. 
In particular, they are proprietary and not available to the public. 
Therefore, it is challenging - if not impossible - to publicly replicate these studies. 

\todo{
Plus/minus type of ratings have been widespread in other sports, mainly basketball, and for the most part they are built using publicly available data (some recent versions of adjusted plus/minus incorporate information from player tracking data). 
The main idea behind these metrics is to allocate the credit of the performance to the different players on the court through a regression. 
Despite the fairly frequent lineup changes in basketball, there are still severe co-linearities in the data since particular players tend to share the court with the same teammates. 
Regularizing the regression has been used to alleviate this problem and improve the predictive power of the obtained ratings. 
Furthermore, there are many different approaches that have been used to regularize the regression, ranging from a typical L1 and L2 regularization to using a Bayesian prior from box score statistics. 
Even though regularization significantly improves the underlying model and ratings obtained, it still requires several observations before they stabilize. 
For example, ESPN does not release its version of plus/minus (RPM) before half of the NBA season is completed. 
The more severe the co-linearity problems (which is the case in soccer), the more challenging will be for the regularization methods to work. 
(\cite{kharrat2020plus}) introduce various versions of the adjusted plus/minus in soccer using both the goals scored during an observation period, as well as the expected goals during the same period. 
The authors use ridge regularizer. 
However, as (\cite{matano2018augmenting}) showed the presence of severe colinearities results to the ridge regression shrinking the coefficients and assigning very similar ratings to players that share the pitch very frequently (which is very common in soccer). 
To solve this problem, they proposed an augmented version of the adjusted plus/minus by using a prior for every players rating based on their rating from the FIFA video game (this is the same dataset we are using in our study). 
Essentially, instead of shrinking the coefficients to a common mean (zero), the augmented version of the plus/minus shrinks each coefficient towards the corresponding FIFA video game player rating. 
This approach is similar to the one behind ESPN's RPM for NBA, where each coefficient is shrunk towards a mean obtained from every players box score\footnote{The most recent updated of RPM also includes player tracking data.}(\cite{espn-rpm}). 
}

\todo{Given the challenge in building an adjusted plus/minus metric for a sport like soccer, we decided to take a different approach. 
In particular, in order to step away from the co-linearity problem we substitute every player with their position and  their FIFA rating. 
This means that our regression will be agnostic to the specific players, but it will rather provide us with the expected on pitch contribution of a player of a given position and  FIFA rating. 
As it will become evident when we describe our approach in detail this essentially will provide an estimation of soccer positional values. 
For instance, how much more important are the midfielders compared to the goalkeeper when it comes to winning a game?
Co-linearity is significantly reduced in our specification since different teams will include players in positions with different FIFA ratings. 
In other words, even though the same players still share the pitch with the same teammates for a large fraction of time, the fact that our independent variable is the FIFA rating of each player allows us to pool observations from many different games that increases the variability in the observations and allows us to identify positional values. 
}

In order to achieve this we use data from games from 11 European leagues as well as FIFA ratings for the players that played in these games. 
\todo{To clarify, these are the ratings from the popular EA Sports video game series, FIFA, similar to (\cite{matano2018augmenting}). However, } 
these ratings have been shown to be able to drive real-world soccer analytics studies (\cite{cotta2016using}), they account for a variety of factors (e.g., player aging) and they are easy to obtain\footnote{Data and code are available at: \url{https://github.com/kpelechrinis/eLPAR-soccer}. Data are also available at (\cite{Kaggle}).}. 
Using these ratings we model the final goal differential of a game through a Skellam regression that allows us to estimate the impact of 1 unit of increase of the FIFA rating for a specific position on the probability of winning the game. 
As we will elaborate on later, to avoid any additional data sparsity problems (e.g., very few teams play with a sweeper today), we group positions in the four team lines (attack, midfield, defense and goalkeeping) and use as our model's independent variables the difference on the average rating of the corresponding lines.  
Using this model we can then estimate the {\bf expected} league points added above replacement ({\method}) for every player. 
The emphasis is put on the fact that this is the expected points added from a player, since it is based on a fairly static, usually pre-season\footnote{FIFA ratings change a few times over the course of a season based on the overall player's performance.}, player rating,  and hence, does not capture the exact performance of a player in the games he played.  
However, when we describe our model in detail it should become evident that if these data (i.e., game-level player ratings) are available the exact same framework can be used to evaluate the actual league points added above replacement from every player.  

The contribution of our work is twofold:

\begin{enumerate}
\item We develop a pre-game win probability model for soccer that is accurate and well-calibrated. More importantly it is based on the starting lineups of the two teams and hence, it can account for personnel changes between games. 
\item We develop the expected league points added above replacement ({\method}) metric that can be used to identify positional values in soccer and facilitate various quantitative applications related to monetary player valuation.
\end{enumerate}

The rest of the paper is organized as follows. 
Section \ref{sec:method} describes the data we used as well as the Skellam regression model we developed for the score differential and its evaluation. 
Section \ref{sec:moneyball} further details the development of our expected league points added above replacement using the Skellam regression model. 
In this section we also discuss applications of {\method} related to monetary evaluations of players. 
Finally, Section \ref{sec:discussion} concludes our work, while also discussing future directions for further improvements of our framework. 

\section{Data and Methods}
\label{sec:method}

In this section we will present the data that we used for our analysis, existing modeling approaches for soccer, as well as, the Skellam regression model we used. 
Table \ref{tab:notations} summarizes some of the notations that we are going to use throughout the paper. 

\begin{table}[htbp]
\begin{center}
\begin{tabular}{r c p{7cm} }
\toprule
$X$ & $\triangleq$ & Goals scored by the home team\\
$Y$ & $\triangleq$ & Goals scored by the visiting team\\
$Z$ & $\triangleq$ & $X-Y$\\  
$\player$ & $\triangleq$ & Individual player \\
$\positions$ & $\triangleq$ & On field position \\
$\positionSet$ & $\triangleq$ & Set of all on field positions \\
$\rating_{\player}$ & $\triangleq$ & FIFA rating for player $\player$\\
$\formation$ & $\triangleq$ & On field team formation \\
$\mv_{\player}$ & $\triangleq$ & Market value for player $\player$ \\
$\cost_{\player}$ & $\triangleq$ & Cost per 1 league point paid for player $\player$\\
$\wage_{\player}$ & $\triangleq$ & (Monthly) Wage for player $\player$ \\
\bottomrule
\end{tabular}
\vspace{0.1in}
\caption{Notations used throughout the study.}
\label{tab:notations}
\end{center}
\end{table}

\subsection{Soccer Dataset}
\label{sec:data}

In our study we make use of the Kaggle European Soccer Database (\cite{Kaggle}). 
This dataset includes all the games (21,374 in total) from 11 European leagues\footnote{English Premier League, Bundesliga, Serie A, Scotish Premier League, La Liga, Swiss Super League, Jupiler League, Ligue 1, Eredivisie, Liga Zon Sagres, Ekstraklasa.} between the seasons 2008-09 and 2015-16. 
For every game, information about the final result as well as the starting lineups are provided. 
There is also temporal information on the corresponding players' ratings for the period covered by the data. 
The overall rating $\rating_{\player}$ for player $\player$ takes values between 0 and 100. 
There are 11,060 players in totals and an average of 2 rating readings per season for every player. 
\todo{Furtermore, we collect information about the players' position from the FIFA's rating website (\url{www.sofifa.com})}.

\todo{One of the pieces of information we need for our analysis is the contract value for a player. 
These type data are harder to obtain - especially compared to North American sports leagues operating with a cap system. 
However, we obtained information for all teams in the English Premier League for the 2015-16 season - the last season covered from our data - from \url{Spotrac.com}. 
Spotrac provides the actual contract value for the season for a player, in contrast to other popular services such as Transfermarkt that provide a crowdsourced estimate for a player's market value (\cite{muller2017beyond}). 
We would like to emphasize here that Transfermarkt data can be accurate in terms of quantifying the value of a player, but they do not represent necessarily the actual value of a player's contract. }


\subsection{Modeling Goals in Soccer}

The goals scored in a soccer game have traditionally been described through a Poisson distribution (\cite{lee1997modeling,karlis2000modelling}), while a negative binomial distribution has also been proposed to account for possible over-dispersion in the data (\cite{pollard198569,greenhough2002football}). 
However, the over-dispersion, whenever observed is fairly small and from a practical perspective does not justify the use of the negative binomial for modeling purposes considering the trade-off between complexity of estimating the models and improvement in accuracy (\cite{karlis2000modelling}). 
In our data, we examined the presence of over-dispersion through the Pearson chi-squared dispersion test. 
We performed the test separately for the goals scored from home and away teams and in both cases the dispersion statistic is very close to 1 (1.01 and 1.1 respectively), which allows us to conclude that a Poisson model is a good fit for our data. 


Another important modeling question is the dependency between the two Poisson processes that capture the scoring for the two competing teams. 
In general, the empirical data exhibit a small correlation (usually with an absolute value for the correlation coefficient less than 0.05) between the goals scored by the two competing teams and the use of Bivariate Poisson models has been proposed to deal with this correlation (\cite{karlis2003analysis}). 
Simply put, $(X,Y)\sim BP(\lambda_1, \lambda_2, \lambda_3)$, where: 

\begin{equation}
P(X=x, Y=y) = e^{-(\lambda_1+\lambda_2+\lambda_3)}\dfrac{\lambda_1^x}{x!}\dfrac{\lambda_2^y}{y!} \sum_{k=0}^{\min (x,y)} \binom{x}{k} \binom{y}{k} k! \bigg(\dfrac{\lambda_3}{\lambda_1 \lambda_2}\bigg)^k
\label{eq:bpois}
\end{equation}
The parameter $\lambda_3$ captures the covariance between the two marginal Poisson distributions for $X$ and $Y$, i.e., $\lambda_3 = Cov(X,Y)$. 
In our data, the correlation between the number of goals scored from the home and away team is also small and equal to -0.06. 
While this correlation is small, (\cite{karlis2003analysis}) showed that it can impact the estimation of the probability of a draw. 
However, a major drawback of the \todo{traditional} Bivariate Poisson model is that it can only model data with positive correlations (\cite{karlis2005bivariate}). 
Given that in our dataset the correlation is negative, and hence, the Bivariate Poisson model from Equation (\ref{eq:bpois}) cannot be used.  
\todo{However, (\cite{mchale2007modelling}) showed that one can use copulas to model bivariate count data with negative dependencies. 
While this can be applied to our problem, given that we are interested in modeling win/loss/draw probabilities and rather the exact score of a game, we follow a different and simpler approach. 
More specifically, 
}
we can directly model the difference between the two Poisson processes that describe the goals scored for the two competing teams.

In particular, with $Z$, $X$ and $Y$ being the random variables describing the final score differential, the goals scored from the home team and the goals scored from the away team respectively, we clearly have $Z=X-Y$. 
With $(X,Y)\sim BP(\lambda_1,\lambda_2,\lambda_3)$, $Z$ has the following probability mass function (\cite{skellam1946frequency}):

\begin{equation}
P(z) = e^{\lambda_1 + \lambda_2}\cdot \bigg(\dfrac{\lambda_1}{\lambda_2}\bigg)^{z/2}\cdot I_z(2~ \sqrt[]{\lambda_1\lambda_2})
\label{eq:skellam}
\end{equation}
where $I_r(x)$ is the modified Bessel function. 
Equation (\ref{eq:skellam}) describes a Skellam distribution and clearly shows that the distribution of $Z$ does not depend on the correlation between the two Poisson distributions $X$ and $Y$. 
In fact, Equation (\ref{eq:skellam}) is exactly the same as the distribution of the difference of two independent Poisson variables (\cite{skellam1946frequency}). 
Therefore, we can directly model the goal differential without having to explicitly model the covariance. 

\subsection{Skellam Regression Model}
\label{sec:skellam_reg}

Our objective is to quantify the value of different positions in soccer. 
This problem translates to identifying how an one-unit increase in the rating of a player's position impacts the probability of his team winning. 
For instance, if we substitute our current striker who has a FIFA rating of 79, with a new striker with a FIFA rating of 80, how do our chances of winning alter? 
Once we have this information we can obtain for every player an expected league points added per game over a reference, i.e., replacement, player (Section \ref{sec:elpar}).  
This can then be used to applications related to the monetary value of a player based on their position and rating (Section \ref{sec:mv}). 

\begin{figure}[t]%
    \centering
    \includegraphics[width=7cm]{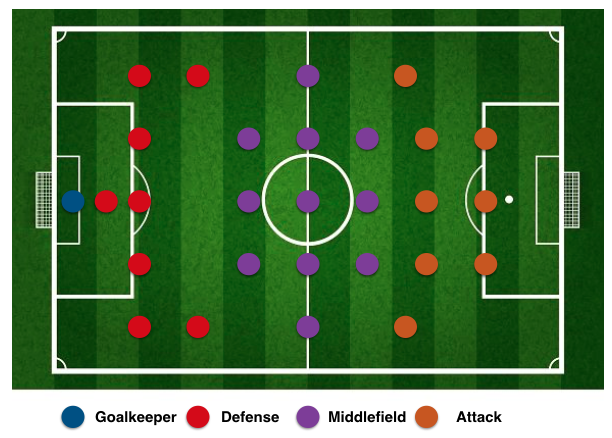} %
    \caption{We grouped player positions to four distinct groups, namely, goalkeeping, attack, midfielders and defense.}%
    \label{fig:positions}%
    \vspace{-0.1in}
\end{figure}

In order to achieve our goal we model the goal differential $Z$ of a game using as our independent variables the player/position ratings of the two teams that compete. 
Hence, our model's dependent variable is the goal differential (home - away) of game $i$, $z_i$, while our independent variables are the positional rating differences of the two teams, $x_{i,\positions}=r_{\player(h,\positions,i)}-r_{\player(a,\positions,i)},~\forall \positions \in \positionSet$, where $r_{\player(h,\positions,i)}$ ($r_{\player(a,\positions,i)}$) is the rating of the home (away) team player that covers position $\positions$ during game $i$ and $\positionSet$ is the set of all soccer positions. 
One of the challenges with this setting is the fact that different teams will use different formations and hence, it can be very often the case that while one team might have 2 center backs and 2 wing backs, the other team might have 3 center backs only in its defensive line. 
This will lead to a situation where the independent variables $x_{i,\positions}$ might not be well-defined. 
While this could potentially be solved by knowing the exact formation of a team (we will elaborate on this later), this is unfortunately a piece of information missing from our data. 
Nevertheless, even this could create data sparsity problems (e.g., formation/player combinations that do not appear often). 
Hence, we merge positions to four groups, namely, attacking line, midfielders, defensive line and goalkeeping. 
Figure \ref{fig:positions} depicts the grouping of the positions we used to the four lines $\linepSet = \{\linep_{\defense},\linep_{\middlefield},\linep_{\attack},\linep_{\gk}\}$. 
Note that this grouping in the four lines has been used in the past when analyzing soccer players as well (\cite{he2015football}). 
The independent variables of our model are then the differences in the average rating of the corresponding lines. 
The interpretation of the model slightly changes now, since the independent variable captures the rating of the whole line as compared to a single position/player. 
Under this setting we fit a Skellam regression for $Z$ through maximum likelihood estimation. 
In particular:

\begin{mdframed}[linecolor=red!60!black,backgroundcolor=gray!20,linewidth=2pt,  topline=true,  rightline=false, leftline=false] 
We model the goal differential $Z_i$ of game $i$ using the following four co-variates:
\begin{itemize}
\item The difference between the average player rating of the defensive line of the two teams $x_{\defense}$
\item The difference between the average player rating of the midfielders of the two teams $x_{\middlefield}$
\item The difference between the average player rating of the attacking line of the two teams $x_{\attack}$
\item The difference between the goalkeeper's rating of the two teams $x_{\gk}$

The random variable $Z$ follows a Skellam distribution, where its parameters depend on the model's covariates $\mathbf{x} = (x_{\defense},x_{\middlefield},x_{\attack},x_{\gk})$:

\begin{eqnarray}
Z \sim Skellam(\lambda_1,\lambda_2)\\
\log(\lambda_1) = \mathbf{b}_1^T \cdot \mathbf{x} \\
\log(\lambda_2) = \mathbf{b}_2^T \cdot \mathbf{x}
\end{eqnarray}
\end{itemize}

\end{mdframed}

Table \ref{tab:skellam_reg} shows the regression coefficients. 
It is interesting to note that the coefficients for the two parameters are fairly symmetric. 
$\lambda_1$ and $\lambda_2$ can be thought of as the mean of the Poisson distributions describing the home and visiting team respectively and hence, a positive relationship between an independent variable and the average goals scored for one team corresponds - to an equally strong - negative relationship between the same variable and the average goals scored for the opposing team. 
An additional thing to note is that an increase on the average rating of any line of a team contributes positively to the team's chances of winning (as one might have expected).  
Finally, having the distribution for the random variable $Z$, we can estimate the win, loss home probability, as well as, the draw probability as:
$\Pr[Home~Win] = \Pr[Z>0]$, $\Pr[Home~Loss] = \Pr[Z<0]$ and $\Pr[Draw] = \Pr[Z=0]$ respectively. 

\begin{table}[ht]\centering

\begin{tabular}{c c c }
\toprule
\textbf{Variable} & \textbf{$\log(\lambda_1)$} & \textbf{$\log(\lambda_2)$} \\ 
\midrule
Intercept         &      0.36776***   &   0.07303***    \\
            &      (0.012)   &      (0.015)  \\
$x_{\defense}$    &  0.01761***             &    -0.02607***     \\
            &       (0.01)        &         (0.002)   \\
$x_{\middlefield}$     &     0.02559***          &      -0.01759***         \\
            &        (0.01)       &             (0.002)    \\
$x_{\attack}$       &       0.00747***&   -0.01095***   \\
            &      (0.001)   &      (0.001)   \\
$x_{\gk}$       &       0.00142&     -0.00313** \\
            &      (0.001)   &     (0.002)      \\
\midrule
 N           &     21,374   &     21,374     \\          
\bottomrule
\addlinespace[1ex]
\multicolumn{3}{l}{\textsuperscript{***}$p<0.01$, 
  \textsuperscript{**}$p<0.05$, 
  \textsuperscript{*}$p<0.1$}
\end{tabular}
\caption{Skellam regression coefficients (standard errors in the parantheses). }
\label{tab:skellam_reg}
\end{table}

\begin{figure}%
    \centering
    \includegraphics[width=6.5cm]{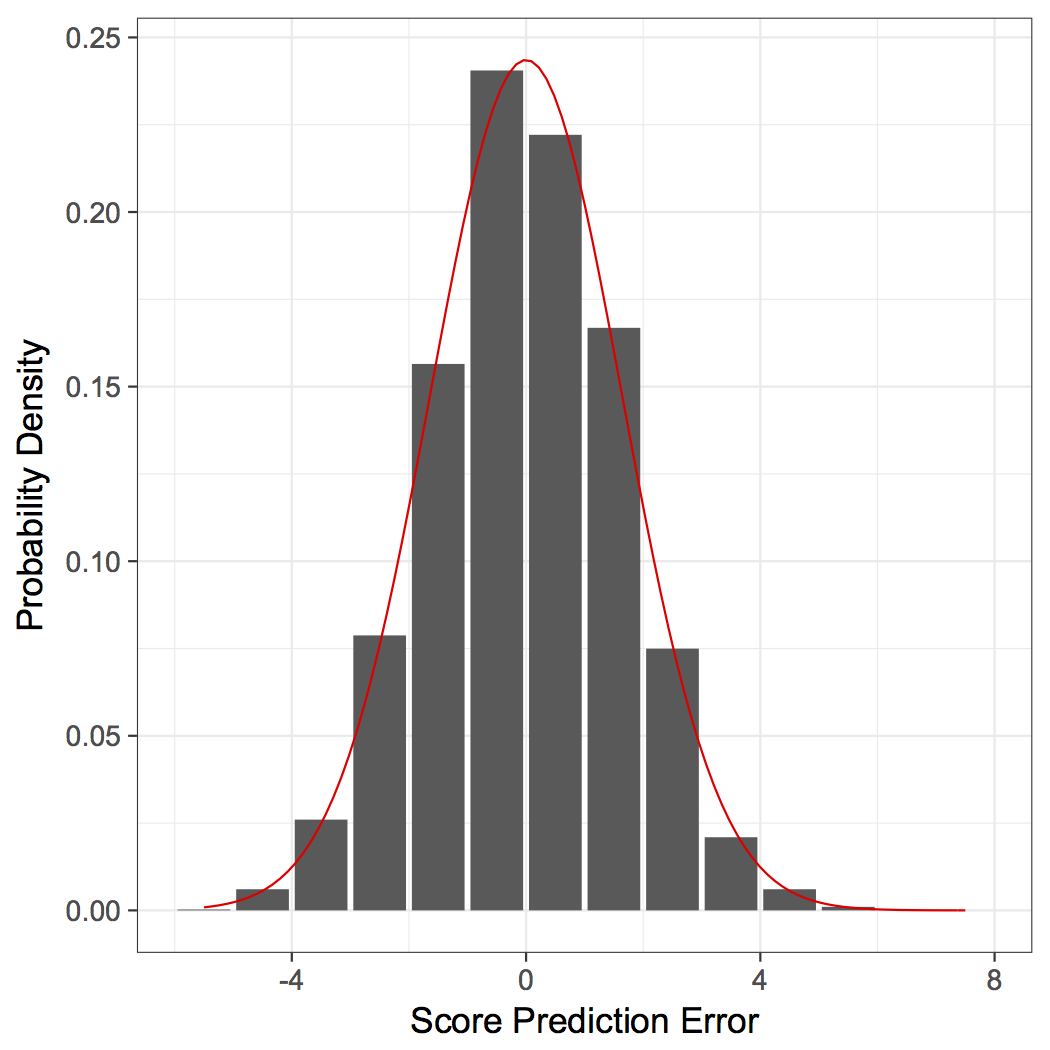} %
    \caption{Our model is accurate in predicting the score differential}%
    \label{fig:model_eval}%
    \vspace{-0.1in}
\end{figure}

Before using the model for estimating the expected league points added above replacement for each player, we examine how good the model is in terms of actually predicting the score differential and the win/draw/loss probabilities. 
We use an 80-20 split for training and testing of the model. 
We begin our evaluation by calculating the difference between the goal differential predicted by our model and the actual goal differential of the game (\cite{10.2307/2684286}). 
Figure \ref{fig:model_eval} presents the distribution of this difference and as we can see it is centered around 0, while the standard deviation is equal to 1.6 goals. 
Furthermore, a chi-squared test cannot reject the hypothesis that the distribution is normal with mean equal to 0 and a standard deviation of 1.6.

However, we would like to emphasize here that the most important aspect of the model is the probability output rather the accuracy of predicting the game outcome. 
Inherently game outcomes include uncertainty and we want our model's probability output to capture this. 
For instance, let us consider two models, $M_1$ and $M_2$ that both predict the home team to win (i.e., a home team win is the most probable among the three possible outcomes). 
$M_1$ assigns a home win probability of 0.4, while $M_2$ assigns a home win probability of 0.7. 
Assuming that the home team wins both have the same accuracy, however it should be clear that they cannot be both accurate in terms of the true home win probability. 
For developing a metric that captures the contribution of a player to his team's win chances, we need a model that provides us with accurate win/loss/draw probabilities. 
As we will see in Section \ref{sec:elpar} we will use the changes in these probabilities to calculate an expected league points added for every player based on their position and rating. 
Hence, we need to evaluate how accurate and well-calibrated these probabilities are. 
This can be evaluated through the \todonew{Brier score and the } probability calibration curves (\todo{\cite{niculescu2005predicting,weisheimer2014reliability,boshnakov2017bivariate}}). 

\todonew{
The Brier score is essentially the mean squared error for the probabilistic predictions. 
In particularly, it is defined from the following equation: 

\begin{equation}
    \bs = \dfrac{1}{N} \sum_{i=1}^N \sum_{j=1}^R (p_{ij} - o_{ij})^2
    \label{eq:brier}
\end{equation}
where $N$ is the total number of observations, $p_{ij}$ is the probability assigned to outcome $j$ for data point $i$ and $o_{ij}$ is a binary indicator that is one if the outcome of data point $i$ is $j$ and zero otherwise. 
Brier score takes values between 0 and 1 and since it is essentially a cost function, a model with lower Brier score is preferable. 
Typically the Brier score of a model is compared to that of a {\em climatology} or baseline model, which assigns to each outcome the base rate probability. 
Specifically, in our dataset home teams win 46\% of the games, visiting teams 29\% of the games and 25\% of the games end with a tie. 
If we assign this probability to each outcome for every game, we can obtain the baseline Brier score. 
For our model, the out-of-sample Brier score is 0.58, while that of the baseline model is 0.65. 
}

\begin{figure}%
    \centering
    \includegraphics[width=8.5cm]{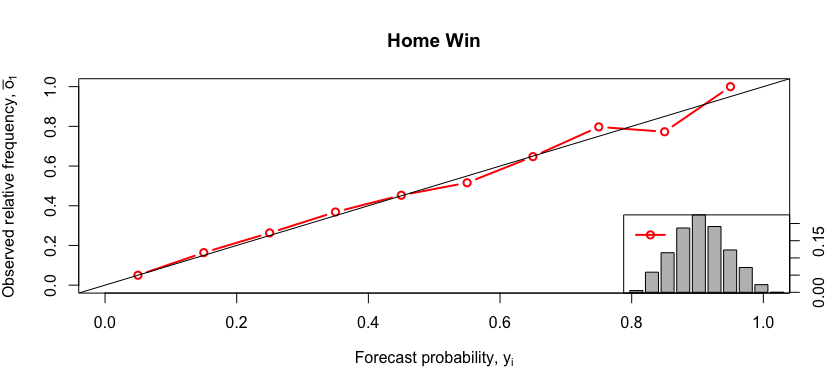} %
    \includegraphics[width=8.5cm]{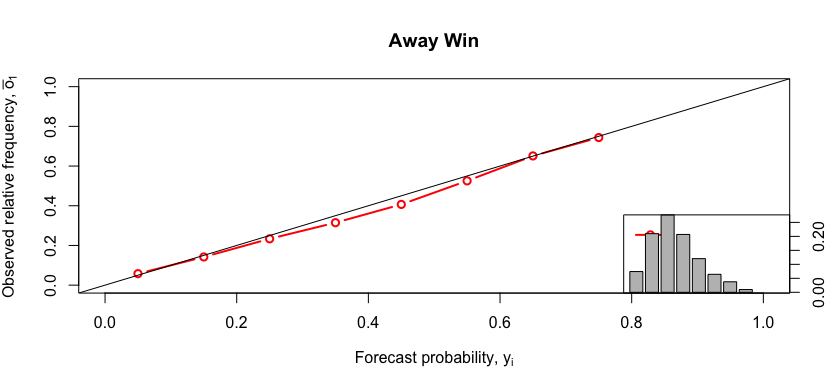}
    \includegraphics[width=8.5cm]{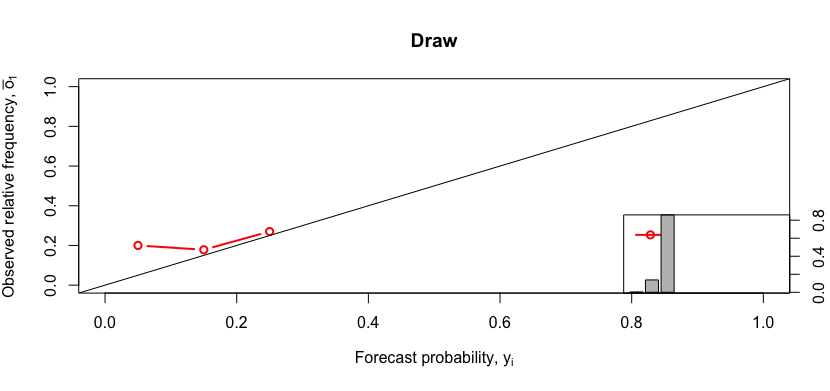}
    \caption{Our model provides well-calibrated probabilities. }%
    \label{fig:model_eval_calibration}%
    \vspace{-0.1in}
\end{figure}

Furthermore, a calibration curve presents on the horizontal axis the predicted probability and on the vertical axis the observed probability. 
More specifically, in order to build the probability calibration curve of a binary classifier we group the test data based on the predicted probability $\pi_{pred}$ of belonging to class ``1''. 
Then for each of these groups we calculate the fraction of the test data points that were indeed of class ``1'', which is the observed probability $\pi_{obs}$. 
Ideally we should have $\pi_{pred}=\pi_{obs}$.
\todonew{Figure \ref{fig:model_eval_calibration} presents the probability calibration curves for our Skellam regression model. 
Given that we have 3 possible results (i.e., home team win, away team win and draw), we present three curves.
The $x$-axis presents the predicted probability for each event, while the $y$-axis is the observed probability. 
In particular we quantize the data in bins of 0.1 probability range, and for all the games within each bin we calculate the fraction of games for which the home team won/away team won/draw, and this is the observed probability. 
The inset in each curve is the distribution of the predicted probabilities in the various bins used for the calibration curve. 
}
To reiterate, we would like to have these two numbers being equal. 
Indeed, as we can see for all 3 events the probability output of our model is very accurate, that is, all lines are practically on top of the $y=x$ line. 
It is interesting to note, that our model does not provide a draw probability higher than 30\% for any of the games in the test set, possibly due to the fact that the base rate for draws in the whole dataset is about 25\%. 
\todonew{Furthermore,the deviation of the draw curve from the $y=x$ line for small probabilities, is most probably due to the fact that there are very few data points in this bin according to the distribution shown in the inset. 
}

\todonew{
\subsubsection{Alternative model specification}
\label{sec:alternative}

One can think of alternative model specifications, and in particularly, models that combine the FIFA ratings of the different lines to create different features for the Skellam regression. 
For example, one natural choice is instead of comparing the ratings of the same lines for the two teams, to compare lines that are interacting on the pitch (e.g., the defensive line of a team and the offensive line of the opponent). 
We built a Skellam regression model with the following features: 

\begin{itemize}
    \item The difference between the ratings of the home attacking line and the visiting defending line
    \item The difference between the ratings of the home middlefield line and the visiting middlefield line
    \item The difference between the rating of the home defending line and the away attacking line
    \item The difference between the home and visiting goalkeeper ratings 

\end{itemize}

Evaluating this model on the same out-of-sample data, we obtain a slightly higher Brier score, 0.59. 
However, the performance of these two models is for all practical purposes identical on this dataset. 
Hence, for the rest of this work we will use our original specification. 

}

\section{eLPAR and Market Value}
\label{sec:moneyball}

We begin by defining the notion of a replacement player and developing {\method}. 
We also show how we can use {\method} to analyze the monetary value (player salaries) of players based on their expected contribution on the pitch. 

\subsection{Replacement Player and Expected League Points Added}
\label{sec:elpar}

The notion of replacement player was popularized by (\cite{woolner2002understanding}) who developed the Value Over Replacement Player (VORP) metric for baseball. 
The high level idea is that player talent comes at different levels. 
For instance, there are superstar players, average players and subpar player talent. 
These different levels come in different proportions within the pool of players, with superstars being a scarcity, while subpar players (what Woolner termed replacement players) being a commodity. 
This essentially means that a team needs to spend a lot of money if it wants to acquire a superstar, while \todo{a replacement player can be acquired for much cheaper (e.g., minimum salary). 
Since a replacement player can be thought of as a {\em minimum salary} player}, a good way to evaluate (and consequently estimate a market value for) a player is to estimate the (expected) contribution in wins, points etc. that he/she offers above a replacement player. 
One of the main contributions of Woolner's work is to show that average players have value (\cite{vorp}).  
Hence, if we were to use the average player as our reference for evaluating talent, we would fail to recognize the value of average playing time. 
Nevertheless, replacement level, even though it is important for assigning economic value to a player, it is a less concrete mathematical concept. 
\todo{(\cite{vorp2}) through his analysis identified that the replacement level has a rating approximately 80\% of the positional average player rating for most of the positions. 
Given that this result is tied to baseball, we are taking a different approach in defining replacement level. 
In particular, according to the abstract definition provided above, a replacement level player is one that can be acquired at minimal cost. 
Therefore, a direct way of defining replacement level is by sorting the contract values of the players in each position, and taking the average rating of the bottom 10th percentile. 
This is essentially saying that the replacement level for each position is defined by the {\em cheapest} players that one can acquire in that position. 
Using the contract values from the EPL 2015-16 season, the replacement levels for the various positions were: 68.3 (goalkeeper), 64.4 (defense), 64.5 (middlefield) and 67.5 (attack). 
}
So the question now becomes how are we going to estimate the expected league points added above replacement ($\method$) given the model from Section \ref{sec:skellam_reg} and the replacements levels of each line. 
First let us define $\method$ more concretely:  

\begin{mdframed}[linecolor=red!60!black,backgroundcolor=gray!20,linewidth=2pt,  topline=true,  rightline=false, leftline=false] 
Consider a game between teams with only replacement players. 
Player $\player$ substitutes a replacement player in the lineup. 
$\method_{\player}$ describes how many league points (win=3 points, draw = 1 point, loss = 0 points) player $\player$ is expected to add for his team. 
\end{mdframed}

Based on the above definition, $\method_{\player}$ can be calculated by estimating the change in the win/draw/loss probability after substituting a replacement player with $\player$. 
However, the win probability model aforementioned does not consider individual players but rather lines. 
Therefore, in order to estimate the expected points to be added by inserting player $\player$ in the lineup we have to consider the formation used by the team. 
For example, a defender substituting a replacement player in a 5-3-2 formation will add a different value of expected points as compared to a formation with only 3 center-backs in the defensive line. 
Therefore, in order to estimate $\method_{\player}$ we need to specify the formation we are referring to. 
Had the formation been available in our dataset we could have built a multilevel model, where each combination of position and formation would have had their own coefficients\footnote{And in this case we would also be able to analyze better the impact of positions within a line (e.g., value of RB/LB compared to CB).}. 
Nevertheless, since this is not available our model captures the formation-average value of each line. 
In particular, $\method_{\player}$ for player $\player$ with rating $r_{\player}$ can be calculated as following:

\begin{enumerate} 
\item Calculate the increase in the average rating of the line $\linep \in \linepSet$ when $\player$ substituted the replacement player based on, $r_{\player}$, formation $\formation$ and the replacement player rating for the line, $r_{replacement,\formation,\linep}$
\item Calculate, using the win probability model above, the change in the win, loss and draw probability ($\delta P_w$, $\delta P_d$ and $\delta P_l$ respectively)
\item Calculate $\method_{\player}(\formation)$ as: 
\begin{equation}
\method_{\player}(\formation) = 3\cdot \delta P_w + 1\cdot \delta P_d
\label{eq:elpar}
\end{equation}
\end{enumerate}

\begin{figure*}%
    \centering
    \includegraphics[width=6.8cm]{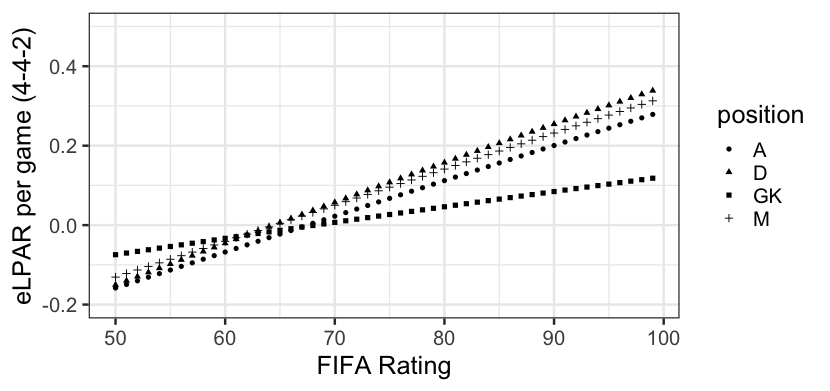} %
    \includegraphics[width=6.8cm]{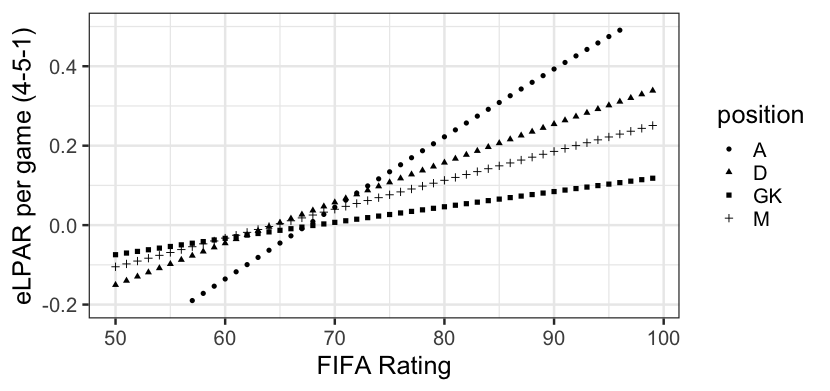} %
    \includegraphics[width=6.8cm]{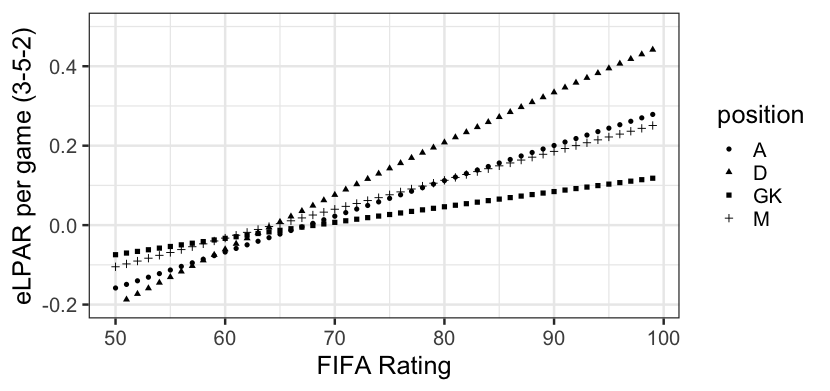} %
    \includegraphics[width=6.8cm]{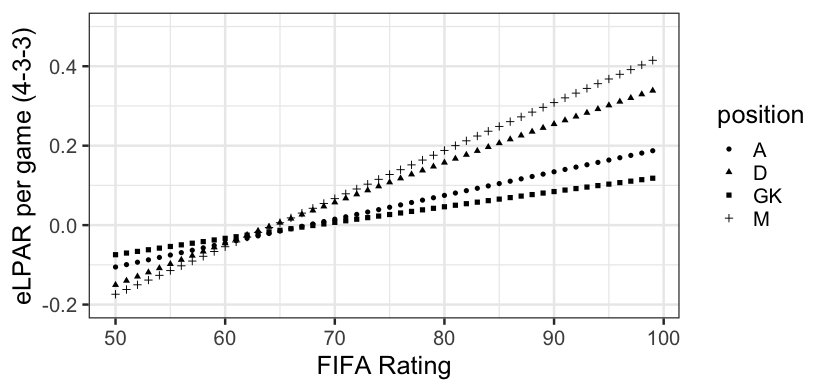} %
    \caption{\todo{Expected league points added above replacement for different formations, player ratings and positions.}}%
    \label{fig:elpar_formations_sub}%
\end{figure*}

It should be evident that by definition a replacement player has $\method = 0$ - regardless of the formation - while if a player has rating better than a replacement, his $\method$ will be positive. 
However, the actual value and how it compares to players playing in different positions will depend on the formation. 
In Figure \ref{fig:elpar_formations_sub} we present the expected league points added per game for players with different ratings (ranging from 50 to 99) and for different formations. 
While there are several different formations that a team can use, we chose 4 of the most often used ones. 

\begin{figure}%
    \centering
    \includegraphics[width=10cm]{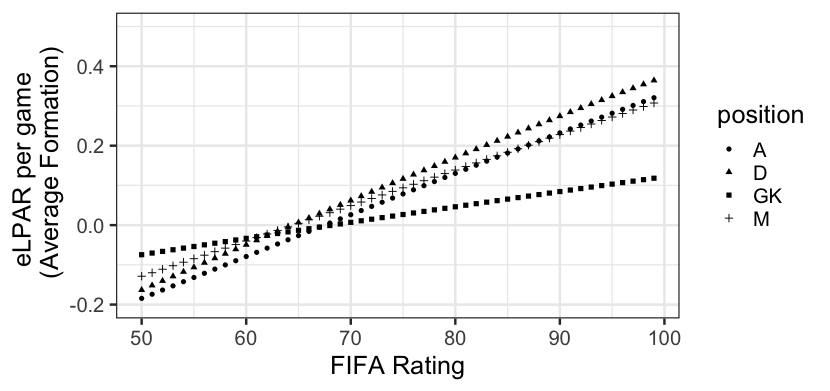} %
    \caption{\todo{Expected league points added above replacement averaging over the four most popular formations for different player ratings and positions.}}%
    \label{fig:elpar_formations}%
\end{figure}
 

One common pattern in all of the formations presented is the fact that for a given player rating goal keepers provide the smallest expected league points above replacement - which is in line with other studies/reports for the value of goal keepers in today's soccer (\cite{Economist-gk}). 
\todo{
Furthermore, one thing to keep in mind when interpreting these results, is that the expected contributions of a player depend mainly on two things (with respect to our formulation of {\method}), namely, the value of the line the player plays in (Table \ref{tab:skellam_reg}) and the formation itself. 
The latter dictates the number of players that are in a line, and hence, the impact that a player's rating will have on the average rating of the line. 
For example, in a 4-5-1 formation there is a single attacker, and hence, substituting a striker with rating $r_1$ for a striker with rating $r_2$, while change the line's average rating by $r_2-r_1$. 
However, for a 4-4-2 formation, the same substitution will have a smaller impact, as it will only change the line's average rating by $0.5\cdot (r_2-r_1)$. 
}
So a 4-5-1 system benefits more from an attacker with a rating of 90 as compared to a defender with the same rating, while in a 3-5-2 formation the opposite is true. 
\todo{While this reliance on the formation might seem as a constraint, one possible usage for a team manager is tailoring the estimation of a potential acquisition target player's {\method} to the dominant formation the manager plans to use.}
To reiterate this is an expected value added, i.e., it is not based on the actual performance of a player but rather on static ratings for a player. 
Given that teams play different formations over different games (or even during the same game after in-game adjustments), a more detailed calculation of $\method$ would include the fraction of total playing time spent by each player on a specific formation. 
With $T$ being the total number of minutes played by $\player$, and $t_{\formation}$ the total minutes he played in formation $\formation$, we have:

\begin{equation}
\method_{\player} = \dfrac{1}{T}\sum_{\formation} t_{\formation} \cdot\method_{\player}(\formation)
\label{eq:elpar_formation}
\end{equation}

Figure \ref{fig:elpar_formations} presents the average $\method$ for each line and player rating across all the four formations (assuming equal playing time for all formations). 
As we can see for the same player rating, a defender adds more expected league points above replacement, followed very closely by a midfielder with the same rating. 
An attacker with the same rating has a lower {\method}, while a goal keeper (with the same rating) adds the least amount of expected league points.
A team manager can use this information to identify more appropriate targets given the team's style play (formations used) and budget. 
In the following section we will showcase some potential applications tying a player's {\method} with his market value.

\subsection{Positional Value and Player \todo{Salaries}}
\label{sec:mv}

\begin{figure*}[h]%
    \centering
    \includegraphics[width=7.5cm]{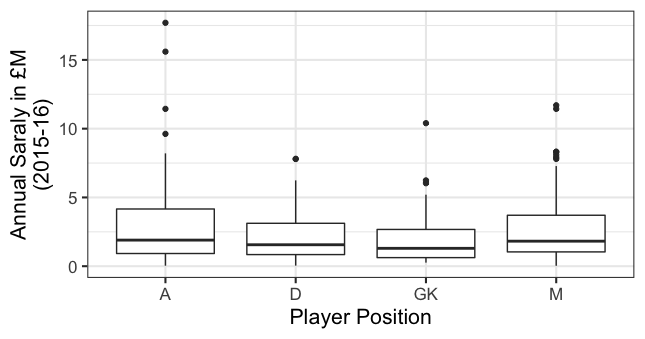}
    \caption{Goalkeepers were among the lowest paid players in the EPL (2015-16 season) in terms of annual salary.}%
    \label{fig:mv}%
\end{figure*}

\begin{figure*}[h]%
    \centering
    \includegraphics[width=7.5cm]{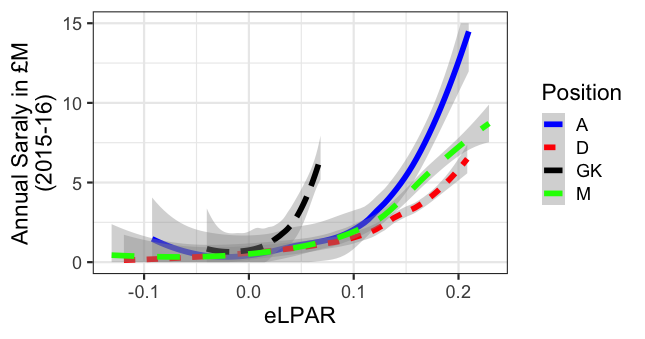} %
    \caption{When looking at the annual salary of a player as a function of their {\method} per game, goalkeepers are paid more compared to other positions that provide the same {\method}. The curves are obtained through a generalized additive model with a cubic spline fitted in the data. }%
    \label{fig:mv2}%
\end{figure*}

\todo{
In this section we will explore how we can use {\method} to connect a player's (expected) performance with their monetary compensation.  
In particular, we are interested in examining whether the  market overvalues specific positions based on the {\method} value they provide. 
Now before we delve into our analysis, we would like to emphasize here that the market for (European) soccer is more complicated in some ways as compared to that of US sports. 
For one there is no salary cap, which makes inter-team comparisons fairly challenging. 
Furthermore, there are many components being involved in a player's monetary compensation, including transfer fees, buyouts, annual salaries etc. that they are not always available in public.  
Also while an annual salary is meant to be compensation for a year, the transfer fee is a ``one-time fee'' that the team has to pay regardless if the player stays for 6 months with the club or for 10 years. 
All these make it hard to analyze the player market in a holistic way\footnote{For instance, Real Madrid after winning the Spanish championship in 2019-20, had to pay an additional 10-15 million pounds to Chelsea for Hazard's transfer. This is part of a clause in the transfer completed the previous summer. Similar details are challenging to be included in our analysis}. 
However, we are going to present here some basic analysis on ways that {\method} can be used in this respect by focusing on annual salaries. 
%
It is - to an extent - realistic to assume that a team will behave the same way when it comes to spending money for a player's salary and paying his transfer fee (or other components of the player's compensation); if a team is willing to pay a high salary for a player, they most probably will also be willing to pay a high transfer fee for him as well. 
Hence, some of the analysis and conclusions {\em might} be transferable to monetary value in general. 

Splitting the players into the four lines, Figure \ref{fig:mv} presents the salaries paid to the players during the 2015-16 season in the English Premier League. 
As we can see goalkeepers are the lowest paid (overall), with defenders coming at a slightly higher price, while  midfielders and attackers are paid gradually higher salaries. 
Of course, these are distributions of the raw data and do not control for other variables such as player quality.  
What we are really interested in is the monetary value that a team pays for 1 expected league point above replacement per player (per game). 
Granted there is a different supply of players in different positions. 
For example, only 8.5\% of the players are goal keepers, as compared to approximately 35\% of defenders\footnote{There is another approximately 35\% of midfielders and 21\% of attackers.}, and hence, one might expect this to factor in when negotiating salaries. 
However, there is also smaller demand for goalkeepers (a team typically has 3 goalkeepers in their active roster), and hence, we expect these two to cancel out to a fairly great extent, at least to an extend that should not over-inflate the market values. 
Given that 1 league point is worth the same in terms of league standings, one would expect that two players with the same {\method} would be paid the same amount regardless of their position. 
However, what we see from  Figure \ref{fig:mv2} is that for players with the same {\method} value there are differences observed in their annual salary depending on their position. 
In particular, goalkeepers have the highest annual salary compared to players of other positions with the same expected on-pitch contribution. 
At the same time, a defender is going to be paid less compared to a forward or a midfielder with the same {\method} (especially for $\method > 0.1$). 

}

To reiterate, the absence of salary cap allows teams to potentially overpay in general in order to bring in the players they want. 
Hence, across teams comparisons are not appropriate, since different teams have different budget and ability to pursue players. 
However, within team comparison of contracts among its players is one way to explore whether teams are being rational in terms of their payroll. 
In particular, we can examine the distribution of a team's total (salary) budget among their players, and investigate whether this is in line with their positional values. 
This analysis will provide us with some relative insights on whether teams spend their player salary budget proportional to the positional and personal on-field value of each player. 

\todo{In particular, we consider all the teams in the 2015-16 EPL season and for each team we estimate the mean {\method} contributed from a player at each line/position and their mean salary.  
We can then obtain the fraction of expected points contributed on average by each line $f_{\method,\linep}$, as well as, the fraction of salary budget $\budget$ allocated to it $f_{\budget,\linep}$. 
We can then obtain the difference between the fraction of resources (salary) allocated to line {\linep} and the fraction of value provided by {\linep}, $f_{\budget,\linep}-f_{\method,\linep}$ (which we term salary-performance deviation $\dev$). 
Table \ref{tab:epl1516} presents our results for each of the teams in our dataset for which we have salary information for at least their starting 11. 
As we can see all of the teams in our dataset underpay their defensive players - when considering purely their {\method}. 
We can further estimate the mean absolute deviation (MAD), as the average of the absolute values for each position. 
This can be thought of as an indicator of how much a team's salary allocation deviates from the expected on-pitch performance. 
The average absolute deviation over all the teams is approximately 0.095, which can roughly be interpreted as a team {\em misallocating} its salaries to the various lines by about 9.5\%. 
Again, we would like to re-iterate that these results come with several caveats. 
For one, we have different percentage of the roster for each team due to data availability. 
Furthermore, particularly for larger rosters, our data includes substitution players that while possibly having a large {\method}, their playing time is limited and hence, their annual salary reflects this aspect and not their ``quality''\footnote{Of course, this should impact all lines the same and eventually we can expect any effects to cancel out.}. 
Essentially while the conclusions themselves should not be taken at face value, the same methodology can provide robust conclusions when fed with complete budget data. 

There is also another interesting thing to note from the  observations above on the positional salaries. 
The positions that appear to be getting ``overpaid'' are the ones that include the fewest players on the pitch at a given time. 
In particular, on the pitch defenders and midfielders outnumber the strikers and goalies by a ratio anywhere from close to 2-to-1 (e.g., at a 4-3-3 formation) up to 4-to-1 (e.g., at a 4-5-1 formation). 
The {\em scarcity heuristic} is a mental shortcut that people use to put a value on an item. 
Using this shortcut can lead to systematic errors and cognitive biases (\cite{lynn1989scarcity}). 
This numeric difference among players of different positions on the pitch (and even on the market) can trigger similar cognitive biases when placing a monetary value on a player. 

}

\begin{table}[!htbp] \centering 
\begin{tabular}{@{\extracolsep{5pt}} ccccc||c} 
\\[-1.8ex]\hline 
\hline \\[-1.8ex] 
Team & $\dev_D$ & $\dev_A$ & $\dev_{GK}$ & $\dev_M$ & MAD \\ 
\hline \\[-1.8ex] 
Arsenal & $$-$0.152$ & $0.101$ & $0.109$ & $$-$0.059$ & $0.105$ \\ 
 Aston Villa & $$-$0.327$ & $0.102$ & $0.196$ & $0.029$ & $0.163$ \\ 
Bournemouth & $$-$0.107$ & $0.091$ & $0.006$ & $0.010$ & $0.053$ \\ 
Chelsea & $$-$0.132$ & $0.095$ & $0.066$ & $$-$0.029$ & $0.081$ \\ 
 Crystal Palace & $$-$0.143$ & $0.038$ & $0.075$ & $0.031$ & $0.071$ \\ 
 Everton & $$-$0.092$ & $0.010$ & $0.064$ & $0.018$ & $0.046$ \\ 
Leicester & $$-$0.263$ & $0.109$ & $0.138$ & $0.016$ & $0.132$ \\ 
Liverpool & $$-$0.146$ & $0.016$ & $0.113$ & $0.017$ & $0.073$ \\ 
Manchester City & $$-$0.119$ & $0.020$ & $0.133$ & $$-$0.034$ & $0.076$ \\ 
 Manchester United & $-0.180$ & $0.064$ & $0.184$ & $-0.068$ & $0.124$ \\ 
 Newcastle & $-0.093$ & $0.051$ & $0.171$ & $-0.129$ & $0.111$ \\ 
 Southampton & $-0.230$ & $0.075$ & $0.229$ & $-0.074$ & $0.152$ \\ 
 Watford & $-0.110$ & $0.029$ & $0.113$ & $-0.033$ & $0.071$ \\ 
\hline \\[-1.8ex] 
\end{tabular} 
  \caption{\todo{Salary-performance deviation for each line and each team in our dataset (EPL 2015-16). Overall, teams underpay defensive players, while paying a premium for strikers and goalies. }} 
  \label{tab:epl1516} 
\end{table}

These results open up interesting questions for soccer clubs when it comes to salary decisions. 
Salary budget is mainly spent for two reasons; (a) to win, as well as, (b) to maximize the monetary return (after all, sports franchises are businesses). 
The premium that clubs are willing to pay an attacker over a defender for the same amount of league points can be seen as an investment. 
These players bring fans in the stadium, increase gate revenue (e.g., through increased ticket prices), bring sponsors, sell club merchandise, etc. 
For example, even though attackers are approximately only 20\% of the players' pool, 60\% of the top-selling jerseys in England during 2018 belonged to attackers (\cite{NBCsports}). 
Therefore, when we discuss the money spent from a team for a player, winning is only one part of the equation. 
While teams with large budget (e.g., Manchester City, Liverpool etc.) might be able to pay premiums as an investment, other teams in the ``middle-of-the-pack'' can achieve significant savings, without compromising their chances of winning. 
In fact, clubs with limited budget can maximize their winning chances, which is an investment as well (winning can bring in revenues that can then be used to acquire better/more popular players leading to a positive feedback loop). 
A club with a fixed budget $\budget$ can distribute it in such a way that maximizes the expected league points {\em bought} (even under positional constraints). 
For instance, with $\budget =  6$ millions and with the need for a center back and a goalkeeper, if we use the median market values for the two positions we should allocate 45\% of the budget (i.e., 2.7 millions) for the goalkeeper and 55\% of the budget for the defender. 
Using the average market value of a player for a given position and rating from our data, this will eventually get us a goalkeeper with an {\method} of 0.045 and a defender with an {\method} of 0.166, for a total of 0.211 {\method} per 90 minutes. 
However, if we allocate 1 million for the goalkeeper and 5 millions for the defender this will get us a total of 0.245 {\method} per 90 minutes, or equivalent this allocation has bought the team 1 expected league point at a 16\% discount as compared to the rest of the market (i.e., with the same amount of money, the team will have obtained 16\% more expected points above replacement per game). 

\begin{figure*}[ht]%
    \centering
    \includegraphics[width=8cm]{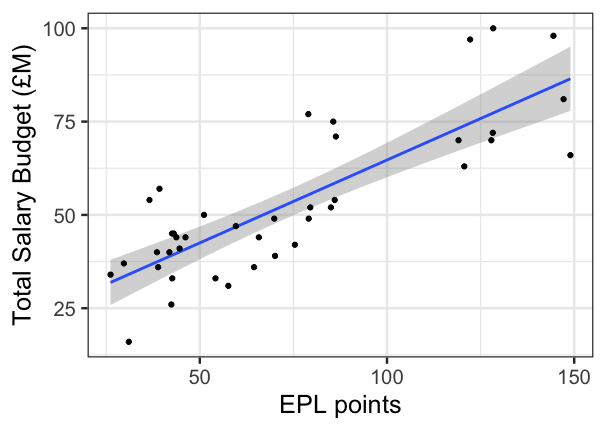}
    \caption{Total annual salary budget and league points for Premier League.}%
    \label{fig:money}%
\end{figure*}

\todo{

We can also examine the relationship between the money a team spends on the players' salaries and the total league points earned during the season. 
For this we can build a simple linear regression, where the independent variable is the total salary budget for a team and the dependent variable is the total points gained from a team. 
Figure \ref{fig:money} presents the data from the 2017-18 and 2018-19 EPL seasons. 
This linear model explains approximately 66\% of the variance for the points earned. 
The slope of the linear fit is 0.45, which translates that 1 million pounds in salary {\em buys} roughly 0.45 Premier League points. 
Obviously this is an oversimplification of the whole processes but we can use this to identify a rough estimate for a {\em fair salary} of a player in the current market.  
For example, for a player $p$ with $\method_p$, who is expected to play $N$ games, a fair annual salary is $\dfrac{N \cdot \method_p}{0.45}$. 
This value is based on the observation that 0.45 EPL points are associated with 1 million pounds in salary spending. 
However, recall from the discussion above (Table \ref{tab:epl1516}) that the current allocation of salaries might not be aligned to the on-pitch expected performance. 
Hence, there is a possibility that with a better allocation of the total salary budget, a team could essentially change the EPL points-1 million pounds ``exchange rate'' to its favor.   
}

\todo{
Finally, we showcase how we can estimate the contribution of a player in a team in terms of final standings league points. 
This can inform applications such end-of-season (team/league) awards. 
In particular, if a player with {\method} $e$ has played $T$ minutes during the season, then his contribution in terms of total league points (above replacement) is $e\cdot\dfrac{T}{90}$. 
In this case, it is more appropriate to use the end-of-season FIFA ratings for the players, since these incorporate their performance through the whole season. 
Table \ref{tab:liverpool} presents the estimation for the latest EPL champions, Liverpool, where we can see that Virgil van Dijk made the largest contribution above replacement level in terms of league points. 
It is interesting to note that if we sum up all the contributions from the players we obtain a total of 74.93 points. 
Liverpool finished its championship campaign with 99 points. 
In order to understand this difference, we need to remember that these total league points presented on the table are  above replacement, i.e., above the points a team consisting only of replacement players would obtain. 
A replacement level team is still expected to earn league points during a season. 
In particular, if we use the Skellam regression model from Section \ref{sec:skellam_reg} and estimate the expected win/draw probability of a replacement team over an average EPL team, we find that this team is expected to win 13.68 points from its home games and 7.1 from its away games, for a total of 20.78 points. 
So eventually, based on {\method}, Liverpool would have been projected to earn 74.93+20.78 = 95.71 points. 
This is very close to its final league points, verifying from a different perspective the usefulness of our model. 
}
  
\begin{table}[!htbp] \centering 
\begin{tabular}{@{\extracolsep{5pt}} ccccc} 
\\[-1.8ex]\hline 
\hline \\[-1.8ex] 
Player & Prosition & Minute Played & FIFA Rating & Total League Points \\ 
\hline \\[-1.8ex] 
Alisson & GK & $2,545$ & $90$ & $2.360$ \\ 
Adrián & GK & $873$ & $76$ & $0.290$ \\ 
\hline
V. van Dijk & D & $3,420$ & $91$ & $9.390$ \\ 
D. Lovren & D & $777$ & $80$ & $1.270$ \\ 
J. Gomez & D & $1,999$ & $82$ & $3.670$ \\ 
A. Robertson & D & $3,113$ & $86$ & $6.990$ \\ 
J. Matip & D & $703$ & $83$ & $1.360$ \\ 
T. Alexander-Arnold & D & $3,176$ & $85$ & $6.810$ \\ 
N. Williams & D & $230$ & $64$ & $0.010$ \\ 
\hline
Fabinho & M & $2,074$ & $86$ & $6.540$ \\ 
G. Wijnaldum & M & $2,948$ & $85$ & $8.870$ \\ 
J. Milner & M & $926$ & $81$ & $2.250$ \\ 
N. Keïta & M & $814$ & $82$ & $2.090$ \\ 
J. Henderson & M & $2,244$ & $85$ & $6.750$ \\ 
A. Oxlade-Chamberlain & M & $1,489$ & $80$ & $3.390$ \\ 
A. Lallana & M & $373$ & $78$ & $0.740$ \\ 
X. Shaqiri & M & $181$ & $81$ & $0.440$ \\ 
C. Jones & M & $122$ & $65$ & $0.010$ \\ 
H. Elliott & M & $6$ & $64$ & $0$ \\ 
\hline
Roberto Firmino & A & $3,003$ & $88$ & $3.640$ \\ 
S. Mané & A & $2,755$ & $90$ & $3.660$ \\ 
Mohamed Salah & A & $2,888$ & $90$ & $3.840$ \\ 
T. Minamino & A & $243$ & $77$ & $0.140$ \\ 
D. Origi & A & $704$ & $78$ & $0.440$ \\ 
\hline \\[-1.8ex] 
\end{tabular} 
  \todo{\caption{Total league points above replacement for every Liverpool player that played in the 2019-20 EPL. } }
  \label{tab:liverpool}
\end{table} 

\section{Conclusions and Discussion}
\label{sec:discussion}

In this work our objective is to build an appropriate model that will allow us to understand positional values in soccer and consequently develop a metric that can provide an estimate for the {\em expected} contribution of a player on the field translated in units that managers and fans associate with (i.e., league points). 
We start by developing a win probability model for soccer games based on the ratings of the four lines of the teams (attack, middlefield, defense and goalkeeper). 
We then translate these positional values to expected league points added above a replacement player ({\method}) considering a team's formations. 
We further showcase how this framework can be useful for financial decisions. 
Our results indicate that specific positions might be over-valued when only considering their contribution to winning the game. 

However, our study is only the first step towards understanding the positional value in soccer. In particular, while our results show that goal keepers might provide the least amount of value, these results are tight to the data we used. Currently we have built a single model for all the leagues in our dataset. Building a separate model for different leagues could reveal differences in the positional value across leagues that might have to do with style of play, strength and skillsets in each league etc. (\cite{Noslo18}). Furthermore, in top-level competition - for which we do not have data (e.g., Champions League) - goal keepers might provide much more value than in the leagues we analyzed, which include both top-tier and lower-tier domestic leagues. 
\todo{In addition, the definition of replacement player requires information for players' annual salaries. Given that we only have access to data from the EPL, our replacement levels are tied to the Premier League. Ideally, each league would have a separate replacement player for each position. While the salary of a player for a specific FIFA rating might not be different across leagues, the overall quality of the players themselves will be different. While the trends will still hold under this assumption, the absolute values can be different. 
However, regardless of these caveats, the analytical framework that we introduced can be replicated on different datasets, with different player ratings/grades etc. }
Our modeling framework can be improved with additional (meta) data. In particular: 

({\bf 1}) Our framework can integrate the actual formation that the teams used. This will allow us to build a multilevel regression model, which will allow us to include covariates for more fine-grained positions (e.g., center back, center midfielder etc.) and obtain a more detailed view of positional value tied to the formation used.

({\bf 2}) We can also include information about substitutions during a game (another piece of information not available to us). This will allow us to (a) obtain a weighted average for the average rating of a line based on the substitutions, and (b) a much more accurate estimate for a player's total playing time.

({\bf 3}) Our current study is based on player ratings obtained from the FIFA video game. While these ratings have been shown to be realistic and accurate with regards to overall player performance during previous seasons, they are not updated at a game-by-game fashion. Therefore, they might not represent the {\em actual} performance of players in each individual game.  These game ratings for example can be composed through appropriate analysis of player tracking data, which can also provide us with information about how much time a combo-player (e.g., a left midfielder who can also play left wing/forward) played at each line, allowing us to obtain a better estimate of the player's contribution in terms of league points above replacement. \todo{Nevertheless, the framework introduced is flexible enough to incorporate any player rating scheme.}

({\bf 4}) We can add interaction terms between the different covariates in the regression model, in order to see how for a example the defensive line interacts with the opposing attack line etc. Furthermore, we can use as our dependent variable the difference in the expected goals, rather than the actual goals scored. Expected goals (xGs) have been shown to be a better predictor of the quality of a team and better predictor of future performance (\cite{StatsBomb-xG}). However, this would also require the availability of player tracking data to estimate the xGs in a game.

Finally, one of the most important contributions of our study is its potential to be applied to other sports that exhibit similar characteristics with soccer that make well-established methods like plus/minus challenging to be applied. For instance, American Football is a good example where colinearities will be severe for a plus/minus approach. Using player ratings from NFL Madden (in a similar way we use player ratings from FIFA), or even player grades from games (e.g., grades from Pro Football Focus) we can evaluate the contribution of 1 unit increase in the Madden rating/PFF grade of a player to the expected points added from a team's play. The latter could be modeled through an expected points model. This could be a significant step towards defining a metric similar to Wins Above replacement for NFL, and finally understanding the contribution of each position in winning.

\bibliographystyle{DeGruyter}
\bibliography{bibliography}

\end{document}